\documentclass[12pt]{iopart}
\usepackage{iopams,epsf,graphicx}
\begin{document}

\title{On the thermoelectricity of correlated electrons in the zero-temperature
limit}

\author{Kamran Behnia\dag\ Didier Jaccard\ddag\ and Jacques
Flouquet\S}

\address{\dag\ Laboratoire de Physique Quantique(CNRS),
 ESPCI, 10 Rue Vauquelin, F-75005 Paris, France}

\address{\ddag\ D\'epartement de la Physique de la Mati\`ere
Condens\'ee, Universit\'e de Gen\`eve, 24 Quai Ernest Ansermet,
Ch-1211, Geneva, Switzerland}

\address{\S\ DRFMC/SPSMS,  Commissariat \`a l'Energie Atomique,
F-38042 Grenoble, France}

\begin{abstract}
The Seebeck coefficient of a metal is expected to display a linear
temperature-dependence in the zero-temperature limit.  To attain
this regime, it is often necessary to cool the system well below
1K. We put under scrutiny the magnitude of this term in different
families of strongly-interacting electronic systems. For a wide
range of compounds (including heavy-fermion, organic and various
oxide families) a remarkable correlation between this term and the
electronic specific heat is found. We argue that a dimensionless
ratio relating these two signatures of mass renormalisation
contains interesting information about the ground state of each
system. The absolute value of this ratio remains close to unity in
a wide range of strongly-correlated electron systems.
\end{abstract}




\section{Introduction}
\paragraph{} Almost two decades ago, Kadowaki and Woods (KW) noticed a universal
correlation between two distinct signatures of electronic
correlation in heavy fermion systems\cite{kadowaki}. In these
compounds, due to a large density of states at Fermi energy, both
the electronic specific heat ($\gamma = C_{el}/T$) and the $T^{2}$
term in the temperature-dependence of the resistivity ($A$ with
$\rho=\rho_{0}+AT^{2}$) are enhanced. KW defined a ratio linking
these two quantities ($A/\gamma^{2}$) and observed that for
various heavy-fermion compounds the magnitude of this ratio is
close to a  value ($a_0 = 1.0 \times 10^{-5} \mu \Omega cm (mol K
/ mJ)^{2}$), which is an order of magnitude higher than the ratio
observed in simple metals\cite{kadowaki,miyake}. More recently,
Tsujii \emph{et al.}\cite{tsujii} have reported that in many
Yb-based compounds with a moderate effective mass the KW ratio is
closer to the value observed in simple metals. It has been argued
that the proportionality $A \propto\gamma^2$ ratio reflects the
large energy dependence of the conduction electron's
self-energy\cite{miyake}.

During the last years, the discovery of
$T^2$ behavior in other remarkable Fermi liquids, such as
SrRu$_{2}$O$_{4}$\cite{maeno}, LiV$_{2}$O$_{4}$\cite{urano},
La$_{1.7}$Sr$_{0.3}$CuO$_{4}$\cite{nakamae}, and
Na$_{x}$CoO$_{2}$\cite{li}, has led to the extension of the KW
plot beyond the heavy fermion compounds. In many of these metallic
oxides, the KW ratio is found to be intriguingly enhanced and the
enhancement has been attributed to unusually large
electron-electron scattering.

A  Fermi liquid is also characterized by the Wilson ratio ($R_{W}=
\frac{\pi^{2}k_{B}^{2}}{3\mu_{B}^{2}} \frac{\chi_{0}}{\gamma}$,
where $k_{B}$ and $\mu_{B}$ are respectively the Boltzmann
constant and the Bohr magneton) which links $\gamma$ to the Pauli
spin susceptibility, $\chi_{0}$\cite{wilson}. This dimensionless
number is equal to unity for free electrons and increases up to
two for a single Kondo impurity of spin 1/2\cite{yamada}. Indeed,
such an enhanced Wilson ratio has been observed in  a variety of
strongly-correlated electronic systems\cite{lee,maeno,nakamae}.

In this paper, we focus on a third ratio connecting two distinct
consequences of strong correlations among electrons. We begin by
recalling that the thermopower of a free electron gas is linear as
a function of temperature. Moreover, the magnitude of the Seebeck
coefficient in this regime is directly proportional to the density
of states at Fermi energy. A dimensionless ratio links the Seebeck
coefficient to the electronic specific heat through the Faraday
number and is equal to $-1$ for free electrons. Our examination of
the available experimental data leads to the intriguing conclusion
that this ratio remains close to $\pm1$ for a wide range of
strongly-interacting electronic systems in spite of their complex
band structure. We will argue that scrutinizing this ratio in a
given compound is a source of insight to the properties of the
ground state.

\section{The Seebeck coefficient of the free electron gas}
In a Boltzmann picture, the thermo-electeric power, also known as
the Seebeck coefficient, is given
by\cite{ashcroft,ziman,abrikosov}:
\begin{equation}\label{1}
   S= - \frac{\pi^{2}}{3} \frac{k_{B}^{2}T} {e}(\frac{\partial\ln\sigma(\epsilon)}{\partial\epsilon})_{\epsilon_{F}}
\end{equation}
Here, $e$ is the elementary charge and $\epsilon_{F}$ the Fermi
energy. The function $\sigma(\epsilon)$, defined as
\cite{ashcroft}:

\begin{equation}\label{2}
 \sigma(\epsilon)= e^{2}\tau(\epsilon) \int\frac{d\mathbf{k}}{4\pi^{3}}\delta(\epsilon-\epsilon(\mathbf{k}))v(\mathbf{k})v(\mathbf{k})
\end{equation}

yields the dc electric conductivity of the system for $\epsilon
=\epsilon_{F}$, where $\mathbf{k}$ is the electron wave-vector and
$\tau(\epsilon)$ is the scattering time. Inserting this expression
into equation 1 yields\cite{ashcroft}:

\begin{equation}\label{3}
   S= -\frac{\pi^{2}}{3} \frac{k_{B}^{2}T} {e}[(\frac{\partial\ln\tau(\epsilon)}{\partial\epsilon})_{\epsilon_{F}}
   +\frac{\int d\mathbf{k}\delta(\epsilon_{F}-\epsilon(\mathbf{k}))\mathbf{M}^{-1}(\mathbf{k})}{\int d\mathbf{k}\delta(\epsilon_{F}-\epsilon(\mathbf{k}))v(\mathbf{k})v(\mathbf{k})}]
   \end{equation}

where $\mathbf{M}^{-1}_{ij}
(=\pm\frac{1}{\hbar^{2}}\frac{\partial^{2}\epsilon(\mathbf{k})}{\partial
k_{i}\partial k_{j}}$) is the inverse of the effective mass
tensor. This expression is a testimony to the difficulty of
interpretation of the temperature-dependence of thermopower.
It contains information on both transport and
thermodynamic properties of the system. The scattering time and
its energy-dependence are only present in the first term of the
right side of the equation. The second term is purely
thermodynamic.

In the simple case of a free electron gas, the second term of Eq.
3 is equal to $\frac{3}{2\epsilon_{F}}$\cite{ashcroft,barnard}.
Moreover, in the zero-energy limit, the energy-dependence of the
scattering time can be expressed as a simple
function\cite{barnard}:

\begin{equation}\label{4}
\tau(\epsilon)=\tau_{0}\epsilon^{\xi}
 \end{equation}

which yields
($\frac{\partial\ln\tau(\epsilon)}{\partial\epsilon})_{\epsilon=\epsilon_{f}}=\frac{\xi}{\epsilon_{F}}$
 for the first term. The simplest case implies an energy-independent
 relaxation time ($\xi=0$). However, alternative cases such as
 ($\xi=-1/2$) are conceivable\cite{barnard,abrikosov}. The latter corresponds to a constant
 mean-free-path, $\ell_{e}$, which implies $\tau=\ell_{e}/v\propto\epsilon^{-1/2}$ \footnote{In the T=0 limit, an
energy-independent $\ell_{e}$, corresponding to the average
distance between two defects is usually taken for granted. This is
thought to be the case even in presence of strong
correlations\cite{millisandlee}.}.

 This leads to a very simple expression for the thermopower of the free electron
 gas:
 \begin{equation}\label{5}
    S= - \frac{\pi^{2}}{3} \frac{k_{B}^{2}}
    {e}\frac{T}{\epsilon_{F}}(\frac{3}{2}+\xi)
\end{equation}

This textbook expression gives a correct estimation of the
magnitude of thermopower in real metals. It also indicates that
whenever the Fermi energy is replaced by a different and smaller
energy scale, the Seebeck coefficient is expected to increase. The
Fermi energy is related to the carrier concentration $n$ and to
the density of states , $N (\epsilon)$. For free electrons, the
link is given by $N (\epsilon_{F}) =3 n /(2\epsilon_{F})$. Using
this expression, Eq. 5 can be written as:
 \begin{equation}\label{6}
    S=  - \frac{\pi^{2}}{3} \frac{k_{B}^{2}T} {e} \frac{N(\epsilon_{F})}{n}
    (1+\frac{2\xi}{3})
\end{equation}

This equation is strikingly similar to the familiar expression for
the electronic specific heat of free
electrons\cite{ashcroft,ziman,abrikosov}:

\begin{equation}\label{7}
   C_{el}= \frac{\pi^{2}}{3} k_{B}^{2}T
   N(\epsilon_{F})
\end{equation}

In this regime, as Ziman has put it\cite{ziman}, thermopower
probes the specific heat per electron. In other words (and
assuming $\xi$=0): $S = C_{el}/ne$, where the unities are V/K for
$S$, J/Km$^{3}$ for $C_{el}$ and m$^{-3}$ for $n$. However, in
order to compare different compounds, it is common to express
$\gamma = C_{el}/T$ in J/K mol units. Therefore in order to focus
on the $S/C_{el}$ ratio, let us define the dimensionless quantity:

\begin{equation}\label{8}
  q = \frac{S}{T} \frac{N_{Av} e}{\gamma}
\end{equation}

where $N_{Av}$ is the Avogadro number. The constant $N_{Av} e =
9.6 \times 10^{5}$ C/mol is also called the Faraday number. For a
gas of free electrons with $\xi$=0 (the simplest case), $q$ is
equal to -1. In the case of an energy-independent mean-free-path,
implying $\xi=-1/2$, $q$ becomes equal to -2/3. Now, if one
imagines to replace the free electrons by free holes (that is to
assume a hollow spherical Fermi Surface) then $q$ would become
equal to +1 and to +2/3.

Note that the conversion factor assumes that there is one
itinerant electron per formula unit which is often (but not
always) the case.  Whenever the density of carriers is
lower(higher) than 1 $e^{-}$/f.u., the absolute magnitude of $q$
is expected to be proportionally larger (smaller) than unity.

Now we turn our attention to the real metals.

\section{Thermoelectricity in real metals}

At a first glance, the relevance of this picture for a
\emph{quantitative} description of thermopower in real metals is
desperate. Even in alkali metals which present quasi-spherical
Fermi surfaces, the temperature-dependence of the Seebeck
coefficient is not linear and in the case of lithium, it is
unexpectedly positive (at least down to the lowest temperatures
investigated)\cite{barnard}. There are a number of well-known
reasons behind this inadequacy.

First of all, a thermal gradient produces a lattice heat current
in addition to the electronic one. Due to electron-phonon
coupling, this leads to an additional contribution to thermopower
dubbed ``phonon drag''\cite{barnard}, which adds up to the
``diffusion thermopower''. The latter is the signal generated by
the diffusive movement of electrons in the absence of the phononic
current. Phonon drag dominates the temperature-dependence of many
metals in a wide temperature range. [An analogous magnon drag
phenomenon occurs in magnetically-ordered metals.] We recall that
the phonon-drag term is proportional to the lattice specific heat
and the latter varies as T$^{3}$ at low temperatures. Therefore,
it does not contribute to an eventually linear Seebeck coefficient
at very low temperatures and does not constitute a complication in
the T=0 limit.

Even the diffusion thermopower of real metals cannot be reduced to
the simple picture of the previous section. Since there are
different types of scattering centers interacting with various
types of carriers, the deconvolution of different contributions is
most often an impossible task. The total thermopower is expected
to be a weighted sum of different contributions. For example, the
Nordheim-Gorter rule, which corresponds to the Matthiessen rule
for resistivity, treats the case of a one-band metal in presence
of several type of scatterers.  According to this rule,
$S=\frac{\sum\rho_{i}S_{i}}{\sum\rho_{i}}$, where the index $i$
designates  distinct contributions to resistivity, $\rho_{i}$ and
thermopower, $S_{i}$\cite{barnard}. In the case of several type of
carriers, one expects each contribution, S$_{j}$, to be weighted
by the respective conductivity, $\sigma_{j}$. A combination of the
two situations should occur in real multi-band metals
\cite{fletcher}. An obvious obstacle for the application of the
free-electron-gas picture (even at T=0) to a multi-band metal
appears: for each band, thermopower S$_{j}$ can be positive or
negative but the sign of corresponding electric conductivity [and
specific heat] is always positive. Therefore, in principle, the
absolute value of the weighted sum which yields the overall
thermopower could be considerably reduced compared to what is
expected for a one-band metal.

The Mott formula for transition metals\cite{mott,colquitt} is a celebrated
milestone in the understanding of thermoelectricity in multi-band
metals. In this two-band picture, light electrons of the band
associated with the $s$-orbital coexist with the heavier ones of
the $d$ band. The dominant mechanism is the scattering of the
light electrons from the wider [$s$] to the narrower [$d$] band,
due to the larger density of states in the latter. This leads to
an additional scattering rate which is proportional to the density
of state of the $d$ band: $\frac{1}{\tau}\propto
N_{d}(\epsilon_{F})$. As a result of this,  thermopower presents a
component proportional to
($-\frac{1}{N_{d}(\epsilon)}\frac{\partial N_{d}( \epsilon
)}{\partial\epsilon})_{\epsilon=\epsilon_{F}}$ which dominates the
free-electron component\cite{barnard}. The Mott formula provides a
qualitative explanation for the enhanced diffusion thermopower in
transition metals. It successfully predicts that the sign of the
additional contribution is different for elements situated in the
beginning and in the end of the series as a result of the
occupancy (or vacancy) of the $d$ orbital.
All these considerations indicate that thermoelectricity in usual
metals (even in reasonably low temperatures) is dominated by many
factors which do not correlate with their specific heat. This may
partly explain one curious anomaly. In spite of being known for
many decades, widely mentioned\cite{ashcroft,ziman,abrikosov} and
commented in detail\cite{barnard}, the free-electron-gas picture
of thermoelectricity has not been \emph{quantitatively} tested.
There is no trace of a systematic investigation of real metals
verifying the simple correlation between specific heat and
thermopower according to Eq. 6-7.

Let us focus on the specific case of heavy fermion compounds
which, due to their giant specific heat, are a natural playground
for this concept.

\section{Thermoelectricity of heavy electrons in the zero-temperature limit}

\begin{figure}
\begin{center}
  \includegraphics{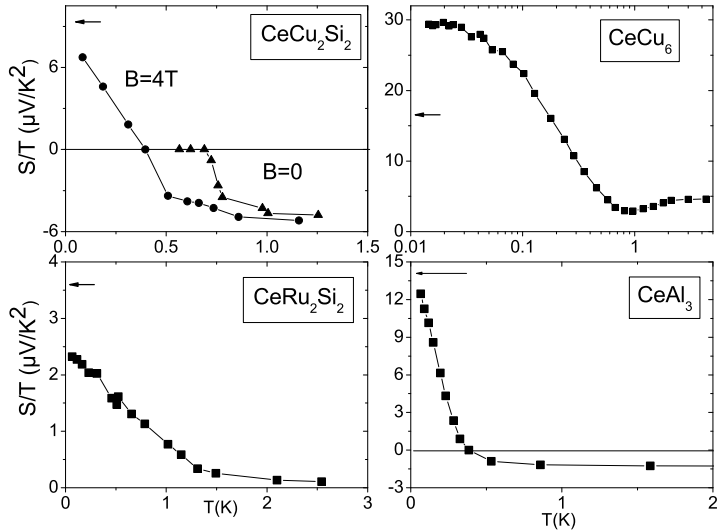}
\caption{$S/T$ as a function of temperature for four different
Ce-based compounds using previously published data by three
different groups\cite{sparn,amato,jaccard1,sato}. In each panel,
the horizontal vector points to the value corresponding to
$\gamma/ (N_{Av} e )$. Note the semi-logarithmic scale in the case
of CeCu$_{6}$.}\label{fig1}
\end{center}
\end{figure}

In heavy fermion compounds(HFC), the effective mass, $m^{*}$ of
quasi-particles is enhanced mainly due to Kondo local fluctuations
around each $f$-electron atom. A new temperature scale ,
T$_{K}\propto 1/m^{*}$ appears which defines a Fermi energy
$\epsilon_{f}= k_{B}T_{K}$ much smaller than in common metals (For
a recent review see \cite{flouquet}).

The investigation of thermoelectricity in HFCs started more than
two decades ago\cite{jaccard7}. An early study on Ce- and Yb-based
compounds displaying a moderate mass enhancement (the so-called
intermediate valence compounds) established a number of features
in qualitative agreement with an extension of the Mott formula to
$f$- electrons\cite{jaccard7,mott2}. Both the large enhancement of
thermopower up to a value close to $k_{B}/e$ and the occurrence of
a maximum at T$_{max}$ corresponding roughly to the bandwidth of
f-electrons (the latter is inversely proportional to $\gamma$) are
compatible with the Mott formula. In many cases, $S$ was found to
remain linear up to a substantial fraction ($\sim1/3$) of
T$_{max}$ and did not show a clear signature of entrance into the
Fermi liquid regime. Moreover, the Mott formula provides a natural
explanation for the positive(negative) sign of thermopower for
Ce(Yb)based compounds in a manner analogous to the case of
transition metals.

During the last two decades, the exploration of numerous HFCs led
to a partial understanding of many features of thermoelectricity
in these compounds. At room temperature, the interplay of
incoherent Kondo scattering with crystal field (CF) effect leads
to a huge $S$ at room temperature. No experimental systematics
appear below the temperature T$_{CF}$ corresponding to the crystal
field energy sacle. Various interpretations have been proposed to
explain why thermopower varies from large and positive (as in the
case of CeCu$_{6}$) to large and negative (as for
CeCu$_{2}$Si$_{2}$) among various compounds. On the other hand,
high pressure studies on cerium
compounds\cite{jaccard5,fierz,jaccard2,link} indicate that under
pressure the positive sign is systematically favored presumably
because the system is driven towards an intermediate valence state
(See\cite{link} for a detailed discussion). On the theoretical
side, in the absence of a microscopic theory of thermoelectricity
in a Kondo lattice, most authors have focused on the
single-impurity case\cite{bickers,newns,houghton}.(For a recent
survey on theory see \cite{zlatic}.)

Nevertheless, the magnitude of the $S/T$ in the zero-temperature
limit and its eventual correlation with $\gamma$ in HFC has not
been a focus of attention. Although such a correlation explicitly
appears in the papers by Read and his
co-workers\cite{newns,houghton}, no experimental study has been
devoted to this issue. Furthermore, for a Kondo impurity of spin
$\frac{1}{2}$ with a complete localisation of the 4$f^1$ charge
(that is  when n$_f =1$), the thermopower is predicted to collapse
at very low temperature. As we will see below, this is not the
case of the cerium Kondo Lattices.

With all these considerations in mind, let us examine the
magnitude of $S/T$ in the zero-temperature limit from an
experimental point of view. In order to address this issue, it is
useful to plot the old thermopower data in a different fashion.

Fig. 1 displays the temperature dependence of $S/T$ using the
previously-published data for four different Ce compounds.  A
polycrystal of CeCu$_{2}$Si$_{2}$ was studied by Sparn \emph{et
al.}\cite{sparn}. At zero field, thermopower remains negative down
to T$_{c}$ ($\sim$ 0.65 K). But, with the application of a
magnetic field and the destruction of superconductivity a positive
$S/T$ emerges. Measurements on a polycrystal of CeAl$_{3}$ was
reported by Jaccard and Flouquet\cite{jaccard1}. Single crystals
of CeRu$_{2}$Si$_{2}$ were studied by Amato \emph{et al.}
\cite{amato} and here we have plotted these data for $S\bot c$. Sato
et \emph{al.} measured a single crystal of CeCu$_{6}$ with the
heat current along the [010] axis\cite{sato}.  As seen in the
figure, in the four cases a finite and positive $S/T$ can be
firmly extracted in the zero-temperature limit. Interestingly, in
all these systems, the value obtained is not very far from the
magnitude of $\gamma/e N_{Av}$. In the case of CeCu$_{2}$Si$_{2}$
and CeAl$_{3}$, the extracted $S/T$ matches $\gamma /N_{Av} e$
within experimental uncertainty ($q \sim 1$). In the two other
compounds the extracted magnitude yield a $q$ close to unity (1.7
for CeCu$_{6}$ and 0.7 for CeRu$_{2}$Si$_{2}$).

The persistent variation of $S/T$ in sub-Kelvin temperature range
indicates that the so-called Fermi-liquid regime in these cases is
established only at very low temperatures. This is backed by a
very careful study of thermopower down to 14 mK in
CeCu$_{6}$\cite{sato}. Indeed, Sato \emph{et al.} reported that
$S/T$ becomes constant only below 30 mK which is also the
temperature associated with the emergence of a purely T$^2$
resistivity\cite{sato}\footnote{The carefully-extracted $A$-term
in CeCu$_6$ \cite{sato}(71 $\mu\Omega cm/K^2$) yields an
anomalously large KW ratio. Interestingly, however, the
discrepancy vanishes if one directly computes the ratio
$A/(\frac{S}{T})^2$ using values obtained below 30 mK. The anomaly
seems to stem from the anisotropy of transport. It is largely
reduced when one compares $\gamma$ with  values of $S/T$ and $A$
averaged along in-plane and out-of-plane directions.}.

\section{A short survey of various families}

\begin{table}
 \centering
\begin{tabular}{|c|c|c|c|c|}
  \hline
 Compound & S/T ($\mu$ V / $K^{2}$) & Remarks & $\gamma$ (mJ /mol $K^{2}$) & $q$ \\

\hline
CeCu$_{2}$Si$_{2}(B=4T)$& 9\cite{sparn}   & polycrystal  & 950\cite{bredl}  &  0.9\\

CeCu$_{6}$& 29\cite{sato} &  along [010]  &1600\cite{lohneysen} & 1.7\\

CeAl$_{3}$& 14\cite{jaccard1} & polycrystal  &1400\cite{jaccard1} & 1..0\\

CeRu$_{2}$Si$_{2}$  &2.4\cite{amato} & in-plane& 350\cite{lacerda} & 0.7\\

CeCoIn$_{5}$ (B=6T) &6\cite{bel} &  in-plane& 650\cite{bianchi} & 0..9\\

CePt$_{2}$Si$_{2}$& 2\cite{bhattacharjee} & along [110]  &130\cite{ayache} & 1.5\\

CeSn$_{3}$ &  0.18\cite{sakurai2}& polycrystal  &    18\cite{ikeda} &   1.0 \\

\hline
CeNiSn & 50\cite{hiess} & polycrystal & 45\cite{izawa} & 107\\
\hline
YbCu$_{4.5}$ &  -7\cite{spendeler}&  polycrystal &635\cite{amato2} &-1.1\\

YbCuAl &  -3.6\cite{jaccard6}&  polycrystal &267\cite{pott} &-1.3\\

YbCu$_{4}$Ag &  -3.6\cite{casanova}&  polycrystal  &200\cite{bauer} &-1.7\\

YbCu$_{2}$Si$_{2}$&  -1\cite{jaccard7,alami-yadri}&  polycrystal &135\cite{sales} &-0.7\\

YbAl$_{3}$&  -0.6\cite{jaccard7}&  polycrystal  &45\cite{walter} &-1.3\\

YblnAu$_{2}$&  -0.75\cite{alami-yadri}& polycrystal & 40\cite{besnus} &-1.8\\

\hline

UPt$_{3}$ & unknown  & none observed\cite{sulpice} & 430\cite{sulpice}  & --\\

UBe$_{13}$(B=7.5T) & -12\cite{jaccard4} & polycrystal & 1100\cite{graf}  & -1.1\\

UNi$_{2}$Al$_{3}$ & 0.24\cite{grauel}  & polycrystal&  120\cite{geibel1} &  0.2\\

UPd$_{2}Al_{3}$ & 0.4\cite{grauel} & S $\bot c $& 150\cite{geibel2} & 0.3\\

URu$_{2}$Si$_{2}$ & -3\cite{sakurai}  & S $\bot c $& 65\cite{maple}  &  -4.5\\

\hline

$\kappa$-(BEDT-TTF)$_{2}$Cu[N(CN)$_{2}]Br$& -0.4\cite{yu} & in-plane & 22\cite{andraka} &  -1.7\\

$\kappa$-(BEDT-TTF)$_{2}$Cu(NSC)$_{2}$& -0.15\cite{sasaki} & in-plane & 25\cite{andraka2} &  -0.6\\

(TMTSF)$_{2}$ClO$_{4}$& unknown & No report found& 11\cite{garoche} & -- \\

\hline
Sr$_{2}$RuO$_{4}$ &  0.3\cite{yoshino} & in-plane& 38\cite{maeno} & 0..8\\

SrRuO$_{3}$ &unknown&  No report found & 30\cite{allen} & --\\

Sr$_{3}$Ru$_{2}$O$_{7}$  &unknown&  No report found & 38\cite{perry} & --\\

SrRhO$_{3}$ & 0.03\cite{yamaura} &polycrystal& 7.6\cite{yamaura2} & 1..3\\

Na$_{x}$CoO$_{2}$ &  0.4\cite{terasaki} &in-plane & 48\cite{ando}  &0..8\\

La$_{1.7}$Sr$_{0.3}$CuO$_{4}$ &  0.18\cite{elizarova} &ceramic& 6.9\cite{nakamae}  &2.5\\

Bi$_{2}$Sr$_{2}$CuO$_{6+\delta}$ &  -0.25\cite{konstantinovic} &ceramic& 8.7\cite{mayama}  &-2.8\\
\hline
NbSe$_{2}$ &  0.3\cite{bel2} &in-plane&17\cite{sanchez} & 1.7\\
Pd & -0.08\cite{fletcher} &polycrystal&  9.5\cite{mizutani} & -0.8\\
Cu & -0.028\cite{rumbo} &along [231] &  1.6\cite{ashcroft} & -1.7\\
constantan (\%43Ni-\%57Cu) & -0.25\cite{barnard} & wire & 27.4\cite{ho} & -0.9\\

\hline
\end{tabular}

\caption{Reported magnitudes of linear thermopower and specific
heat for a number of metals. The significance of the coefficient
$q= \frac{S}{T} \frac{N_{Av} e}{\gamma}$ is discussed in the
text.}\label{T1}
\end{table}

\begin{figure}
\begin{center}
\includegraphics{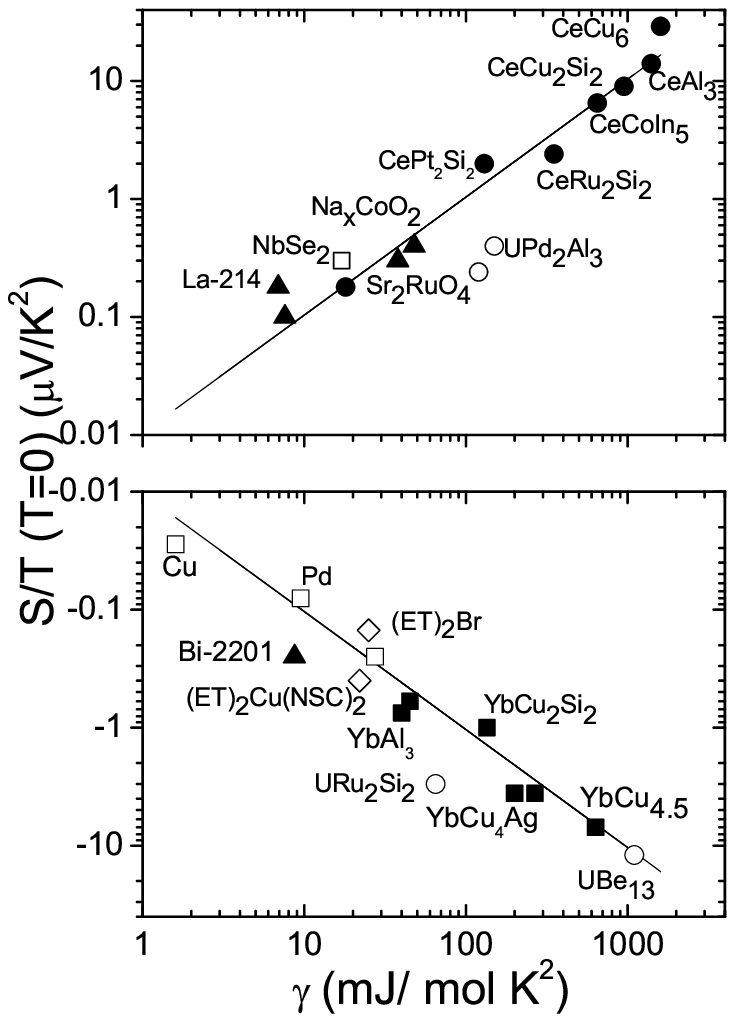}
 \caption{$S/T$ vs. $\gamma$ for the compounds listed
in table 1. Solid circles (squares) represent Ce (Yb)
heavy-fermion systems. Uranium-based compounds are represented by
open circles, metallic oxides by sold triangles, organic
conductors by open diamonds and common metals by open squares. For
some data points, due to the lack of space, the name of the
compound is not explicitly mentioned. See table 1 for the missing
names. The two solid lines represents $\pm\gamma /(e
N_{Av})$.}\label{fig2}
\end{center}
\end{figure}

The specific heat of many remarkable metals is well-documented in
technical literature. This is not, however, the case for
thermoelectric power. In particular, the magnitude of $S/T$ in the
zero-temperature limit is almost never explicitly reported. In
table I, we have compiled the reported data for a number of
compounds. We have tried to restrict ourselves to the cases where
 the extrapolation of data at lowest reported
temperature to T=0 does not appear to produce any significant
change in the sign and/or magnitude of $S/T$. In the case of
low-dimensional systems, we have taken the \emph{in-plane} value.
As seen in the table, in most cases the coefficient $q$ is not
very far from unity. This can also be seen in Fig. 2 which plots
$S/T$ as a function of $\gamma$. Each data point represents a
compound and together they constitute a cloud around a straight
line representing $N_{Av}e/\gamma$. Below, we consider different
families of compounds represented in table I.

\emph{Heavy Fermions:}  In all Ce-based compounds listed in the
table, the ratio $q$ remains close to unity. As $\gamma$ extends
over two orders of magnitude from CeSn$_{3}$ to CeCu$_{6}$, this
correlation between specific heat and thermopower is indeed
remarkable. Note that the sign of thermopower is positive for all
Ce-based compounds.  On the other hand, the thermopower of Yb
compounds which often display a clearly linear
temperature-dependence in a reasonable temperature window is
negative. In all cases the magnitude of a $S/T$ yields $q\sim-1$.

The table also includes CeNiSn, a so-called ``Kondo insulator''.
Given the extremely low carrier density of the system, the very
large magnitude of $q$ ($\sim 106$) is not a surprise. The Hall
data suggests a carrier density of 0.01 $e^{-}/f.u.$ at 5K and
still lower below\cite{takabatake}. The magnitude of $q$ is in
good agreement with this estimation.

The situation is different for the U-based compounds. An early
study of UPt$_{3}$ did not detect a finite $S/T$ at sub-Kelvin
temperatures\cite{sulpice}. [The magnitude of $S/T$ above T$_{c}$
yields $q\sim 0.2-0.3$]. In UBe$_{13}$ thermopower changes
strongly with magnetic field. The largest field applied (7.5T) in
the only reported study\cite{jaccard4} was not enough to destroy
superconductivity. Taking the value of $S/T$ in presence of such a
field at T$\sim$0.8K yields a $q\sim -1$. For the other U-based
compounds of the list, no data is available for low temperatures
and in presence of a magnetic field needed to destroy
superconductivity. In the T=0 limit, the magnitude (and the sign)
of $S/T$ in UPd$_{2}$Al$_{3}$ and UNi$_{2}$Al$_{3}$ could be
somewhat different from what is given in table I which gives the
zero-temperature extrapolation of the data reported for
$T>2K$\cite{grauel}.  In the case of
URu$_{2}$Si$_{2}$\cite{sakurai} there is a simple reason for
expecting a $q$ much larger than unity. Indeed, both Hall effect
measurements\cite{schoenes} and band calculations\cite{yamagami}
indicate that the carrier density at low temperatures is very
small (about 0.05 $e^{-}/U$-atom). Thus the apparently large $q$
($\sim$ 4.5) is a consequence of an enhanced conversion factor
between $\gamma$ and $S/T$. In fact, given such a small carrier
density in URu$_{2}$Si$_{2}$, a $q$ as large as twenty and
sensibly larger than what is given in the table is expected.
Clearly, a fresh look at the thermopower of U-based compounds in
the subkelvin regime would be very useful. Even at this stage,
however, the problem of the 4$f$-electron localisation in uranium
compounds appears to be more complex than in the case of Ce and Yb
compounds. Notably the sign of the thermopower is strikingly
different among different compounds.

\emph{Organic superconductors:} Few studies of thermoelectricity
in organic superconductors are available. The table indicates data
found in literature for two members of the
$\kappa$-(BEDT-TTF)$_{2}$X family of quasi-two-dimensional
superconductors. While $\gamma$ is roughly the same in
$\kappa$-(BEDT-TTF)$_{2}$Cu(NSC)$_2$ and in
$\kappa$-(BEDT-TTF)$_{2}$Cu[N(CN)$_{2}$]Br, two different groups
report sensibly different values of thermopower at the onset of
superconductivity for each compound\cite{yu,sasaki}. In neither
case $S$ is purely linear at  this temperature. The reported
results point to a $q$ in the 0.6 - 1.7 range. Note that the
in-plane thermopower of these compounds is anisotropic and the
values of the table correspond to the [larger] negative ones
attributed to the carriers associated with the
quasi-one-dimensional sheet of the Fermi surface\cite{mori}.
Determination of the magnitude of the low-temperature $S/T$ in the
metallic state of the Bechgaard salts (the (TMTSF)$_{2}$X family)
would be very useful for the purpose of this investigation. The
zero-temperature thermoelectricity of organic conductors with
their simple and well-defined Fermi surfaces appears to be
largely unexplored.

\emph{Metallic oxides:} The table includes a number of metallic
oxides known to be remarkable Fermi liquids. The thermopower of
Sr$_2$RuO$_4$ has been studied down to 4.2 K\cite{yoshino}. It
displays an almost linear temperature dependence over an extended
temperature range. Taking the value of $S/T$ at 4.2K yields
$q=0.8$. We did not find any report on the thermoelectricity of
two other ruthenate compounds displaying a comparable mass
enhancement in their specific heat. It is interesting to observe
that available data for thermopower\cite{terasaki} and specific
heat\cite{ando} of the recently-discovered cobaltite compound
(Na$_{x}$CoO$_{2}$) points also to a $q$ close to unity.  Ando
\emph{et al.}\cite{ando} have already made a qualitative link
between the giant thermopower and the enhanced specific heat in
this case.

Let us underline the interesting case of the heavily overdoped
cuprate La$_{1.7}$Sr$_{0.3}$CuO$_{4}$. At this doping level,
superconductivity is completely absent and resistivity displays a
purely T$^2$ temperature as expected for a Fermi
liquid\cite{nakamae}. Now, a study of thermopower in
La$_{2-x}$Sr$_{x}$CuO$_{4}$ reports that for x=0.3, in contrast
with lower doping levels, thermopower becomes almost linear below
20K with $S/T \sim 0.18\mu V/K^{2}$\cite{elizarova}. This,
combined with $\gamma \sim 6.9mJ/(mol)K^{2}$\cite{nakamae}, yields
$q\sim2.5$. The result is far from anomalous and is to be compared
with 3.3 which is the expected value of $q$ for a system with a
carrier density of 0.3 $e^{-}$ per unit cell. Interestingly, a
linear term, slightly larger and \emph{with an opposite sign}, can
be extracted from the data reported for overdoped Bi-2201 at a
comparable doping level ($p$=0.29). Future studies on single
crystals would be very useful to refine the issue. An intensive
debate on thermopower of the cuprates has focused on the influence
of the doping level on the magnitude of Seebeck coefficient at
high temperatures\cite{obertelli}.

\emph{Common metals:} The extraction of an intrinsic linear
thermopower is particularly difficult in simple elemental metals.
This is due to the small magnitude of thermopower at low
temperatures and its sensitivity to the presence of a small
concentration of impurities. We have found compelling data for
very pure Cu\cite{rumbo} and for hydrogen-free Pd\cite{fletcher}.
The value of $S/T$ has been taken at the lowest reported
temperature (1.5-2K) which is below the last low-temperature
structure in $S$. Interestingly, constantan, a Cu-Ni alloy widely
used as a thermocouple presents a linear $S$ up to room
temperature. Taking this quasi-constant $S/T$ and $\gamma$ yields
a $q$ close to -1. It is tempting to attribute the absence of any
detectable phonon drag in this alloy to presence of strong
disorder which kills electron-phonon coupling. Finally, we have
also included the data from a recent study on the
Charge-Density-Wave compound NbSe$_{2}$\cite{bel2} which present a
positive thermopower and $q\sim 1.7$.

\section{Discussion and unanswered questions}

The principal observation reported in this paper is presented in
Fig.2. Most of the systems considered lie close to the two lines
representing $\pm \frac{\gamma}{N_{Av}e}$. Moreover, in many other
cases which appear not to follow this general trend, the number of
carriers per formula unit gives a satisfactory explanation for the
magnitude of $q$.

Let us stress that, in spite of its conformity to the
free-electron-gas picture, this observation does not lie on a
solid understanding of microscopic properties. Many of the systems
considered here have notoriously complicated Fermi surfaces. In a
naive multi-band picture, the contribution of hole-like and
electron-like carriers would cancel out and lead to a more or less
homogenous distribution of points between the two $\pm
\frac{\gamma}{N_{Av}e}$ lines. Clearly, this is not the case.

One may invoke an inherent asymmetry of mass renormalization
between electrons and holes in each system. Take the case of
cerium and ytterbium compounds. In Ce compounds, the
\emph{occupancy} of \emph{f}-level orbital is expected to lead to
the formation of a narrow band which has a curvature opposite to
the one formed in Yb compounds. Now,  it is a tiny vacancy
($\epsilon$) in the 4$f^1$ content (n$_f=1-\epsilon$) of the 4$f$
shell which is responsible of the Kondo dressing and of the
positive sign of the thermoelectric power in Ce compounds. In the
Yb case, on the other hand, the excess in the 4$f$ content
(respective to the trivalent state Yb$^{+3}$) leads to the
negative sign of $S$. Note that such an explanation is quite
different from the one resulting from the extension of the Mott
formula which also correctly predicts the positive (negative) sign
of thermopower for Ce(Yb) compounds. According to the latter, the
sign of the thermopower is determined by $\frac{\partial
N(\epsilon)}{\partial \epsilon}$ since it affects \emph{the energy
dependence of the scattering rate of the light electrons}. In
other words, the sign of $S$ is imposed by the first term of Eq.3
and not the second which prevails in the simple free-electron-gas
picture. Clearly, a rigorous theoretical investigation of this
issue is required.

Let us also note that within the current resolution, the
experimental data presented in Fig.2 does not allow to detect any
deviation from the general tendency for different families of
correlated-electron systems. This is also remarkable, since large
deviations from the KW value (from $0.04a_{0}$ in several Yb
compounds\cite{tsujii} to $50a_{0}$ in Na$_{x}$CoO2\cite{li} have
been reported. This may not be as surprising as it appears. There
is a fundamental difference between the KW ratio and $q$. While
the former compares the size of \emph{inelastic} electron-electron
scattering with the density of states at Fermi energy, the latter
is a ratio of two zero-energy properties of the system. In this
regard, it is more akin to the Wilson ratio. However, contrary to
the latter, it should mirror those anomalous transport properties
which affect the energy-dependence of the scattering rate.

Finally, we should mention that the observation reported here can
be used as a tool for tracking non-trivial physics associated to
an anomalous value of $q$ at very low (yet finite) temperature.
This is the case of several HF superconductors such as
CeCoIn$_{5}$, UBe$_{13}$ and CeCu$_{2}$Si$_{2}$ at the onset of
superconductivity.

\section{Acknowledgements}
We thank R. Bel and H. Aubin for stimulating discussions.

\end{document}